\documentclass[aip,pof,preprint,amsmath,amssymb]{revtex4-1}
\usepackage{CJK}
\usepackage{dcolumn}
\usepackage{bm}
\usepackage{graphicx}
\usepackage[utf8]{inputenc}
\usepackage[T1]{fontenc}
\usepackage{mathptmx}
\usepackage{siunitx}
\begin{document}

\begin{CJK*}{UTF8}{}
\title{Spontaneous formation of density waves in granular matter under swirling excitation} 



\author{Song-Chuan Zhao \CJKfamily{gbsn}(赵松川)}
\email[]{songchuan.zhao@xjtu.edu.cn}
\affiliation{State Key Laboratory for Strength and Vibration of Mechanical Structures, School of Aerospace Engineering, Xi{'}an Jiaotong University, Xi{'}an 710049, China}

\author{Thorsten P\"{o}schel}%
\affiliation{ 
	Institute for Multiscale Simulation, Friedrich-Alexander-Universit\"{a}t, Cauerstra\ss e 3, 91058 Erlangen,Germany
}%

\date{\today}

\begin{abstract}
We study here the spontaneous clustering  of a submonolayer of grains under horizontal circular shaking. The clustering of grains occurs when increasing the oscillation amplitude beyond a threshold. The dense area travels in a circular fashion at the driving frequency, even exceeds the speed of driving. It turns out that the observed clustering is due to the formation of density wave. The analysis of a phenomenological model shows that the instability of the uniform density profile arises by increasing the oscillation amplitude and captures the non-monotonic dependence of the transition amplitude of the clustering on the global density of the system. Here, the key ingredient is that the velocity of individual grains increases with the local density. The interplay of dissipative particle-particle interaction and the frictional driving of the substrate results into this dependence, which is tested with discrete element method simulations. 
\end{abstract}

\maketitle 
\end{CJK*}


\textit{Introduction} -- Owing to its non-equilibrium nature, granular materials exhibit phenomena of self-organization across orders of magnitude of length scales, from gold panning~\cite{schnautz2005} to astrophysics.~\cite{goldhirsch2003,herminghaus2017} Those phenomena are largely represented by the clustering instability, non-uniform density distribution developing out of an initially homogeneous state. 
Clustering has been observed both in freely cooling granular gas~\cite{goldhirsch1993,Maas2008} and in driven systems~\cite{umbanhowar1996}. Such a collective behavior leads to pattern formation,~\cite{krengel2013} segregation,~\cite{aumaitre2003,schnautz2005,Barker2021} phase separation~\cite{herminghaus2017} and shear banding.~\cite{Kollmer2020} Though clustering of granular matter exhibits some generic features across various systems, to unravel the underlying mechanism one may need to take peculiarities of any given experimental protocol, \textit{e.g.,} the type of energy input, into account.~\cite{cafiero2000} The discovery of new features challenges the existing concepts and theories.~\cite{hummel2016} Unraveling the physics mechanism represents a crossroad of hydrodynamics, nonequilibrium statistical mechanics and the phenomenological theory of pattern formation, which has attracted interest over decades. One may refer to Ref.~\citenum{herminghaus2017} and Ref.~\citenum{aranson2006} and references therein for an overview.   
In this article, we study a submonolayer of beads under horizontal agitations. The constant frictional driving of the substrate distinguishes it from vertically vibrated systems. Strip-like patterns were reported in such systems subjected to a one-dimensional oscillation.~\cite{krengel2013} Under two-dimensional oscillations, a liquid-solid transition was found.~\cite{aumaitre2003} The transition therein was realized by  increasing the global packing density $\phi_{tot}$, while keeping the oscillation amplitude constant. The transition packing density is reduced by larger oscillation amplitude, an example of `freezing by heating'.~\cite{Helbing2000} However, the mechanism of the spontaneous clustering is still unknown, and the motion within the clustering region is not investigated in details. Here, we confirm that, for a given $\phi_{tot}$, the clustering is achieved by increasing the oscillation amplitude. The transition is abrupt and sensitive to the amplitude (see Supplementary Material). The mechanism is explored by analyzing the motion of individual particles, a phenomenological model and DEM Simulations.

\begin{figure}
	\centering
	\includegraphics[width=8.5cm]{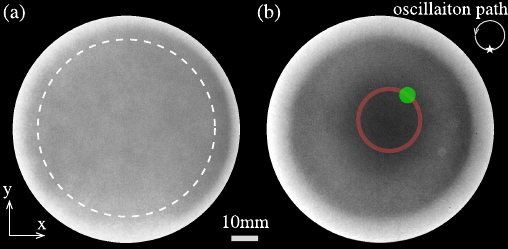}
	\caption{\label{f.img} The average of the experimental images during 10 cycles in the reference frame of the moving container for two oscillation amplitudes: (a) $A=5\si{mm}$ and (b) $A=11\si{mm}$. Within the cylindrical container the background is bright. The darker a region looks, the denser it is. In (a) the dashed circle indicates the region of interest. In (b) dynamic clustering can be observed near the center (Multimedia view). The red circle indicates a stationary path of diameter $22\si{mm}$ {\it in the lab frame of reference}, along which the local density profile in Fig.~\ref{f.DW}a is computed. See the main text for more information. The green solid circle indicates the typical region where individual grains would access during one oscillation cycle. The white circle represents the anti-clockwise oscillation path in scale, whose diameter is 11mm. The position of both displayed images in the oscillation cycle is denoted as $\bigstar$. }
\end{figure}

\textit{Experimental resutls} -- For the system studied here, there are three experimental parameters: the oscillation strength, the global packing density and the  ratio of the grain size to the container size. We first study a reference system specified and later we investigate the influence of various parameters to the system behavior. The submonolayer consists of $N_{tot}=6930$ polydisperse Zirconium Oxide spheres of diameter $ [0.6 \dots 0.8]$ mm with uniform distribution and the mean $d_g=0.7\si{mm}$. The grains located on an acrylic plate are confined by a 3D-printed PLA circular side wall of diameter $D=82\si{mm}$ and height 5\si{mm}. The global packing density is given by the area ratio $\phi_{tot}=N_{tot}(d_g/D)^2$. The inclination of the bottom plate is smaller than 0.02 mm/m. The container is subjected to anti-clockwise circular oscillation in the horizontal plane of frequency $f=5\si{Hz}$. The oscillation amplitude is varied in the range $A=[5 \dots 13] \si{mm}$. Note that here the amplitude represents the diameter of the circular oscillation path (see Fig.~\ref{f.img}b). In this range of agitation and $\phi_{tot}=0.505$, grains rarely jump over each other and, thus, the packing remains two-dimensional. The system is illuminated by a LED panel from the bottom, and the dynamics are captured by a high speed camera (Mikrotron MC1362) at the top at a constant frame rate of 500Hz. The camera is fixed in the laboratory frame of reference. The velocity of particles is obtained by multiplying the distance traveled between consecutive frames by the frame rate, corresponding to a time interval of 0.002\si{\second}. In the following, the analysis is done in this frame for reference.  To avoid the potential influence of the boundary layers, we exclude particles closer than $15d_g$ to the side wall from the analysis (see Fig.~\ref{f.img}a).

Upon oscillation grains roll and slide on the substrate and collide with each other and the side wall. The density distribution changes with the oscillation amplitude. Figure~\ref{f.img} shows the average of images of the system during 10 cycles at two oscillation amplitudes. For $A=5\si{mm}$, the system is homogeneous. For $A=11\si{mm}$, a high density region appears close to the center. The observed clustering transition is very sensitive to the oscillation amplitude. The cluster disappears within 10 cycles after decreasing $A$ from 11mm to 10mm. 
The reversibility of the transition highlights the uniqueness of the clustering in the current work with respect to that in a vertically vibrated monolayer,~\cite{olafsen1998} where hysteresis is observed. Furthermore, in our experiments the dense area moves anti-clockwise, the same direction as the oscillation (Multimedia view). 

This reversible transition is so abrupt that it can be well recognized by naked eyes, which is further confirmed by quantitative measurements of the average and the variance of the local packing density $\phi_{loc}$ in the region of interest. The transition is accompanied with a jump of the average of $\phi_{loc}$, and the variance reaches a peak just below the transition indicating the emergence of unstable small clusters  (see an example in Supplementary Material). The quantitative definitions of $\phi_{loc}$ will be given below.


In Fig.~\ref{f.img}b there are dense areas at the periphery of the packing as well. Those dense boundary layers may be sustained over many cycles and move in a counter-intuitive way, opposite to the swirling motion. This motion mode will be investigated in another work. Nevertheless, the occurrence of a dense area near the periphery is not surprising. The frictional driving of the substrate introduces both linear and angular momentum of grains. The rotational degree of freedom of grains reduces the linear momentum transfer from the substrate. Therefore, the linear speed of grains is 2/7 of that of oscillation,~\cite{kondic1999} ignoring interactions between particles. On one hand, this velocity difference leads to compression on the periphery of the packing via collisions between the particles and the side wall. On the other hand, it always leaves an empty area not containing grains near the side wall, as if the packing only occupies a fraction of the total area (see Fig.~\ref{f.img} for example). In consequence, the packing density in the region of interest, $\phi_{0}$, is typically larger than $\phi_{tot}$. As long as the packing remains two-dimensional, $\phi_{0}$ can be estimated via $\phi_{tot}/\phi_{0}=(1-\tfrac{5}{7}\tfrac{A}{D})^2$.  However, this effect could not explain the observed spontaneous clustering in the central area, for instance, the sensitivity of the clustering to $A$ (see Supplementary Material). To understand the mechanism of the clustering we study the dynamics of individual particles.


\begin{figure}
	\centerline{\includegraphics[height=5cm]{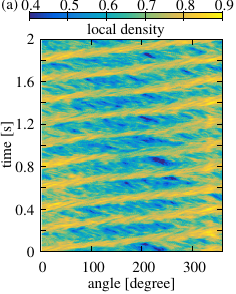}~~~~~
		\includegraphics[height=5cm]{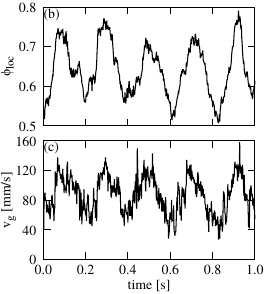}}
	\caption{\label{f.DW} (a) Local density profile along a circular path versus time for $A=11\si{mm}$ and $\phi_{tot}=0.505$, where clustering close to the center is observed. The circular path used here is highlighted in red in Fig.~\ref{f.img}b. The exact position of the path is given in the main text. For the same experiment, the local density $\phi_{loc}$ and the velocity of a grain $v_g$ on the circular path are plotted versus time in (b) and (c) respectively. See main text for the definition of $\phi_{loc}$.}
\end{figure}

\begin{figure}
	\centering
	\includegraphics[width=8cm]{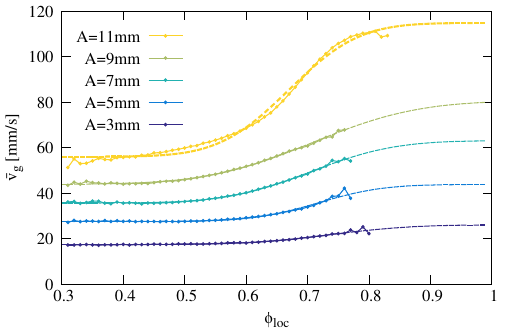}
	\caption{\label{f.vg}The presumed value of the velocity of grains, $\bar{v}_g$, increases with the local density, $\phi_{loc}$, for various oscillation amplitude $A$. $A=11\si{mm}$ triggers the clustering/density wave close to the center of the packing. The dashed lines are fits of the error function to the data. The global packing density here is $\phi_{tot}=0.505$.}
\end{figure}

To visualize the motion of the dense area, we select a circular path in the lab frame and calculate the local density profile along this path. The path is concentric with the moving line of the center of the bottom plate, but has a larger diameter of 22 mm. The local density $\phi_{loc}$ is defined for individual grains in a circular neighborhood with a diameter of  $5d_g$~
\footnote{The conclusion is unchanged for the diameter of the neighborhood region between 3$d_g$ and $7d_g$.}
. The upper bound of $\phi_{loc}$ is $\phi^*\approx 0.9$ corresponding to the hexagonal packing. $\phi_{loc}$ along this circular path is plotted versus time in Fig.~\ref{f.DW}a. At a given time there is a jump of $\phi_{loc}$ in space. The maximum of $\phi_{loc}$ travels along the selected path at the same frequency as the driving (5 Hz). For lower $A$ the density pattern disappears (See Fig.S1 for an example). It is noticeable that the length of the selected path is $22\pi$ mm, twice the swirling motion of the container $\pi A=11\pi\si{mm}$. In other words, the motion of $\phi_{loc}$ covers twice the distance of the oscillation itself during one period. As explained above, due to the rotational degree of freedom, the grains move slower than the container. This implies that the observed motion is the propagation of density waves. 
Meanwhile the trajectory of individual grains is still confined to a region much smaller than the size of the oscillation path. For comparison, the region in which a grain on the circular path moves during one cycle is highlighted by the green solid circle in Fig.~\ref{f.img}b. The local density around this grain and its velocity are plotted in Fig.~\ref{f.DW}b-c. Its neighborhood experiences periodic compression and dilation. The velocity of the grain, $v_g$, follows the same periodic pattern and shows a positive correlation with $\phi_{loc}$, \textit{i.e.,} grains in denser regions tend to move faster. Figure~\ref{f.vg} shows the velocity of the particle as a function of $\phi_{loc}$ averaged over the region of interest and time (not just of those on the selected circular path). For the experiment in Fig.~\ref{f.DW} ($A=11\si{mm}$) $\bar{v}_g$ saturates for low and high $\phi_{loc}$, but increases steeply between $\phi_{loc}=0.6$ and $\phi_{loc}=0.8$. Note that a similar dependence of $\bar{v}_g$ on $\phi_{loc}$ already appears for lower oscillation amplitude ($A<11\si{mm}$), where no clustering/density wave is observed. At low $\phi_{loc}$, $\bar{v}_g$ is close to $2/7A\pi f$, as that of a sphere rolling on the oscillating substrate without sliding.~\cite{kondic1999} Henceforth $\bar{v}_g$ is referred to as the \emph{presumed velocity} of grains.

\textit{Stability analysis} -- The dependence of $\bar{v}_g$ on $\phi_{loc}$ reveals the mechanism leading to the clustering. For a given $A$, consider the continuity equation:
\begin{equation}
	\frac{\partial\phi_{loc}}{\partial t}+\nabla\cdot(\phi_{loc}\vec{v})=0.
	\label{eq.continue}
\end{equation}
If the velocity of particles tends to relax towards the local presumed velocity $\vec{V}(\phi_{loc})$, it is readily to show that an increasing function of $\vec{V}$ on $\phi_{loc}$ would promote the formation of shock waves of $\phi_{loc}$.~\cite{Liu1987,Friedlander2002} However, the density wave is only observed for $A\geq 11\si{mm}$, which suggests that the collisions between grains introduce an equivalent term of pressure sustaining the homogeneous state. Therefore, the equation for the velocity field can be written as
\begin{equation}
	\frac{\partial \vec{v}}{\partial t}+\vec{v}\cdot\nabla\vec{v} = \frac{\vec{V}_g(\phi_{loc})-\vec{v}}{\tau}-\frac{1}{\phi_{loc}}\nabla p(\phi_{loc}).
	\label{eq.vel}
\end{equation}
The first term at the right hand side is the tendency of the local velocity to match the presumed velocity, $\vec{V}_g$, where $\tau$ is the time scale of the relaxation of $\vec{v}$ towards $\vec{V}_g$. As the interactions between particles are via contacts/collisions, $\tau$ is defined by the collision time scale. It can be seen in Fig.~\ref{f.DW}(c) that $\tau$ is much smaller than the oscillation period.
\footnote{Otherwise $v_g$ would display an apparent delay relative to $\phi_{loc}$, e.g., reaching its maximum at a later time.} The second term represents the gradient of the pressure, $p$. This term prevent particles staying in the denser region, where they bounce away from each other via frequent collisions.~\footnote{Unlike in the free-cooling granular gas, the particles here are subjected to constant driving from the substrate.} Note that both $\vec{V}_g$ and $p$ are functions of $\phi_{loc}$ and $A$. The presumed velocity, $\vec{V}_g$, largely follows the direction of the oscillation. Therefore, we only consider the flow in the oscillation direction, and Eq.~\ref{eq.vel} is reduced to a scalar equation. The one-dimensional equation corresponding to Eq.~\ref{eq.vel} was derived in the context of vehicular traffic models in Ref.~\citenum{Kurtze1995} 

The model embodied in Eqs.~\ref{eq.continue} and \ref{eq.vel} admits a steady-state solution representing the uniform flow ($\phi_{loc}=\phi_0$ and $v=V_g(\phi_0)$). Note that we use $\phi_0$ instead of $\phi_{tot}$ for $\phi_{loc}$. Such a homogeneous flow is stable against density perturbations, provided~\cite{Kurtze1995}
\begin{equation}
	(\phi_{loc}V'_g)^2<p'.
	\label{eq.stable}
\end{equation}
$V'_g$ and $p'$ are the derivatives of $V_g$ and $p$ with respect to $\phi_{loc}$ for a given $A$. If condition~\ref{eq.stable} is violated, the destiny disturbance grows and travels at a higher velocity than $V_g(\phi_{0})$ corresponding to homogeneous flow. Error function is fitted on the measured $\bar{v}_g(\phi_{loc},A)$ (see Fig.~\ref{f.vg}), and it is used as a substitute for $V_g$. Thus, the derivative, $\bar{v}'_g$, has Gaussian-like peaks. The pressure from collisions is estimated by
\begin{equation}
	p=c_0\overline{\delta{v}_g^2}\frac{d_g^2}{\bar{d}(\bar{d}-d^*)}=c_0\overline{\delta{v}_g^2}f(\phi_{loc}).
	\label{eq.pressure}
\end{equation}
with the dimensionless parameter $c_0\approx 2$ (see Supplementary Material) whose value is later determined by comparing with the experimental observation. For individual grains, the velocity fluctuation $\delta v_g$ in its neighborhood is extracted. Similar to $\bar{v}_g$, $\overline{\delta{v}_g^2}(\phi_{loc},A)$ is the average of $\delta v^2_g$ over the region of interest and time. $\bar{d}=d_g/\sqrt{\phi_{loc}}$ represents the average distance between grains for a given $\phi_{loc}$, and $d^*=d_g/\sqrt{\phi^*}$ corresponds to the hexagonal packing. The fraction on the right hand side of Eq.~\ref{eq.pressure} is purely geometrical and is referred to as $f(\phi_{loc})$ in the following. $f(\phi_{loc})$ increases with $\phi_{loc}$ and diverges when approaching $\phi_{loc}=\phi^*$. In contrast, though increasing with $A$, $\overline{\delta v_g^2}$ is largely constant in range of $\phi_{loc}\in [0.3,0.8]$ for a given $A$ (see Supplementary Material). Therefore, the variant of $p$ is dominated by $f(\phi_{loc})$, and its derivative is approximated by $p'\approx c_0\overline{\delta{v}^2_g}f'(\phi_{loc})$.

\begin{figure}
	\centering
	\includegraphics[width=8cm]{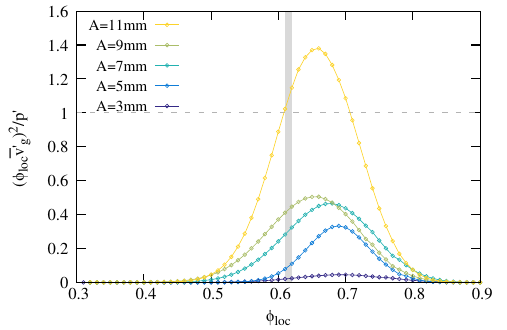}
	\caption{\label{f.model} The ratio of the two terms in Eq.~\ref{eq.stable} are plotted. The region where the ratio is larger than 1 indicates the violation of Eq.~\ref{eq.stable}. The vertical gray bar indicates the density in the region of interest, $\phi_0$, for the oscillation amplitude without clustering. The global packing density is $\phi_{tot}=0.505$.}
\end{figure}

Figure~\ref{f.model} illustrates the condition Eq.~\ref{eq.stable}. The packing density in the region of interest for $A<11\si{mm}$ is indicated by a gray bar ($\phi_0\in [0.61,0.62]$). $c_0=1.8$ is chosen such that Eq.~\ref{eq.stable} is just violated for $A=11\si{mm}$ at $\phi_0$, but not for smaller oscillation amplitude.  Though both $p'$ (or $\delta v_g^2$) and $\bar{v}'_g$ increase with $A$, the relative increase of the latter is more significant. In consequence, the instability is triggered by increasing $A$ beyond the transition amplitude, $A_c$. This corresponds to the observed `freezing by heating'. 

Figure~\ref{f.model} provides further insight. The instability associated with the violation of Eq.~\ref{eq.stable} would only be initialized for packings of intermediate densities, where the ratio $(\phi_{loc}\bar{v}'_g)^2/p'$ displays a peak. The peak shape leads to a non-monotonic dependence of the transition amplitude, $A_c$, on the global packing density of the system, $\phi_{tot}$. Imagine that $\phi_{tot}$ is increased from $0.505$ (analyzed so far) towards $\phi^*$. Before reaching the peak of $(\phi_{loc}\bar{v}'_g)^2/p'$ the transition amplitude, $A_c$, would reduce. However, it would raise quickly again when $\phi_{tot}$ is increased beyond the peak. $p'$ diverges at $\phi_{tot}=\phi^*$, and so does $A_c$. 

\begin{figure}
	\centerline{\includegraphics[width=8cm]{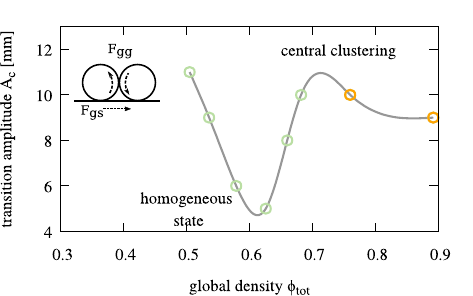}}
	\caption{\label{f.tr} The transition amplitude, $A_c$, triggering the clustering is a non-monotonic function of the global packing density $\phi_{tot}$. The lowest $\phi_{tot}$ of the spontaneous clustering is 0.51 for the tested range of $A$. The last two points where the 2D packing scenario breaks down are highlighted by orange. The solid line serves as guide for the eye. Inset: a sketch of the situation of rotational frustration.}
\end{figure} 

The above hypothesis is examined in experiments. The function $A_c(\phi_{tot})$ is plotted in Fig.~\ref{f.tr}. The clustering at all $\phi_{tot}$ is observed for 100 cycles to confirm the transition. {The typical size of the clustering increases from $30d_g$ at $\phi_{tot}=0.505$ till $60d_g$ at $\phi_{tot}=0.64$ and saturates beyond that.~\footnote{The boundary of a cluster is identified first by binarizing the map of $\phi_{loc}$ and locating the largest connected object in the binary map. The cluster size is then estimated from the largest dimension of the cluster. The midpoint of the rapid decay of $\phi_{loc}$ across the front of the density wave (see Fig.~\ref{f.DW}a) is chosen as the threshold for binarization. } 
 The lowest density where spontaneous clustering is observed is $\phi_{tot}\approx 0.505$. The minimum of $A_c$ is reached for $\phi_{tot}= 0.64$. We confirm that the variation of $A_c$ observed in Fig.~\ref{f.tr} does not alter the monotonic increase of $\phi_{0}$ along with $\phi_{tot}$, except for the two highest $\phi_{tot}$. The shape of the function $A_c(\phi_{tot})$ is in qualitative agreement with the model. It needs to be emphasized that the measured $\bar{v}_g(A,\phi_{loc})$ in Fig.~\ref{f.vg} depends on $\phi_{tot}$. In consequence, for $\phi_{tot}>0.505$ the ratio of $(\phi_{loc}\bar{v}'_g)^2/p'$ is quantitatively different from that shown in Fig.~\ref{f.model} (see Supplementary Material). This does not alter the peak shape of $\bar{v}'_g$ that leads to the non-monotonic change of $A_c$ with $\phi_{tot}$. However, the secondary decrease of $A_c$ for $\phi_{tot}\gtrsim 0.7$ is not expected in the model, since the divergence of $p$ at high $\phi_{tot}$ is supposed to suppress any density wave formation. In practice, when the pressure rising from the collision between grains exceeds that resulting form gravity, grains jump over each other, and the divergence of $p$ and the 2D packing scenario used so far breaks down. For dense packings ($\phi_{tot}\gtrsim 0.7$) we observed such behavior near the wall, and the packing density close to the center is decreased, which allows the formation of clusters. On the other hand, for very small $A$, the highest $\phi_{tot}$ packing could maintain the two-dimension configuration and appears as a `full' cluster. In this case, the whole packing rotates clockwise, opposite to the swirling motion. This distinctive motion mode will be further investigated in a following work.


\textit{Discussions} -- We have elaborated the mechanism of clustering by the instability analysis of the uniform flow in a phenomenological model (Eqs.~\ref{eq.continue} and \ref{eq.vel}). The key ingredient promoting the clustering is the dependence of the presumed velocity of grains on the the local density. Why does such a dependence exist? We believe that it is a consequence of the interplay between the friction between grains and that between grains and the substrate. $\mu_{gg}$ and $\mu_{gs}$ denote the friction coefficients of the former and the latter respectively, and $F_{gg}$ and $F_{gs}$ are the corresponding friction forces. Upon agitation $F_{gs}$ accelerates not only the linear momentum but also the angular momentum of grains. As discussed above, without interactions, individual grains would reach a linear speed of 2/7 of the oscillation speed, rolling without sliding.\cite{kondic1999} 
It also implies that a grain would reach a larger linear speed, if its rolling speed is reduced. Consider now the collision of two spherical grains rolling in the same direction (inset of Fig.~\ref{f.tr}). $F_{gg}$ counteracts the rolling of grains. The frustration of rotation effectively enhances the action of $F_{gs}$ on linear momentum transfer. $F_{gg}$ is proportional to kinetic pressure $p$, and $p$ increases with $\phi_{loc}$ and $A$  (Eq.~\ref{eq.pressure}). Therefore, grains tend to move faster in dense areas and/or under stronger oscillations. These arguments lead to the features of the function $\bar{v}_g(\phi_{loc},A)$ shown in Fig.~\ref{f.vg}, where $\bar{v}_g$ increases with both $\phi_{loc}$ and $A$. In order to further support this line of arguments discrete element method simulations are performed using \textsc{liggghts}.~\cite{kloss2012} For given $\mu_{gs}$ and $A$, by decreasing $\mu_{gg}$, the dependence of $v_g$ on $\phi_{loc}$ and the clustering are both suppressed. Meanwhile, as observed in experiments, reducing $A$ takes a similar effects on clustering, which is confirmed for a few combinations of oscillation frequency and amplitude. The transition amplitude $A_c$ decreases with $f$, and a constant critical oscillation strength $A_cf^2$ can be identified for relatively low frequencies. However, at high frequency $A_c$ deviates from this trend (see Supplementary Material). Simulations quantitatively consistent with experiments will be appreciated to reveal the relative significance of parameters, where a careful exam on the force model, such as the coupling between rolling and sliding~\cite{kondic1999}, is required.

The clustering phenomenon reported here could be reproduced in experiments with polydisperse grains of diameter 0.8-1 mm, various surface types (smooth and rough glass beads), aspheric grains (e.g., millet seeds) and in a square container (see Supplementary Material). Therefore, the principle of clustering is robust for the explored parameter range. For even larger aspect ratio of $d_g/D$ a solid-like cluster and its reptation motion mode were reported,~\cite{Scherer1996,Scherer2000} which is different from the phenomenology here in both the global density distribution (Fig.~\ref{f.img}b) and the motion of the cluster (Fig.~\ref{f.DW}a). Moreover, the 'liquid-solid' transition was reported to be independent of the oscillation frequency.~\cite{aumaitre2003} Therefore, further exploration of the parameter space is necessary to reach a complete quantitative understanding of the clustering in swirling granular matter, its growth/coarsening and the potential relation to segregation.~\cite{aumaitre2003} In particular, the abrupt increase of ${\bar{v}'}_g$ across $A_c$ (cf. Fig.~\ref{f.vg} and Fig.S3) and its dependence on $\phi_{tot}$ indicate certain non-local effect, which increases with $\phi_{tot}$ and helps reducing $A_c$ for $\phi_{tot}$ between 0.50 and 0.64. An appropriate hydrodynamic treatment of this term is anticipated in the future research.

\section*{Supplementary Material}
See the Supplementary Material for details of DEM simulations and snapshots of clustering in other experimental configurations.

%

This work is supported by German Science Foundation.

The  data  that  support  the  findings  of  this  study  are  available from the corresponding authors upon reasonable request.
\bibliography{density_wave}
\end{document}